\documentclass[5p]{elsarticle}

\pdfoutput=1

\usepackage{natbib}
\setcitestyle{authoryear,longnamesfirst}
\usepackage{siunitx}
\usepackage{graphicx}
\usepackage{amsmath}
\usepackage{xspace}
\usepackage{algorithm}
\usepackage{algpseudocode}
\usepackage{geometry}
\usepackage[toc,page]{appendix}
\usepackage{caption}
\usepackage{subcaption}
\usepackage{multirow}
\usepackage[acronym]{glossaries-extra}
\usepackage{tikz}
\usepackage{pgfplots}
\usepackage{datetime}

\pgfplotsset{compat=1.17}
\usepackage{tikzscale}
\usepgfplotslibrary{fillbetween}
\usetikzlibrary{calc}

\newcommand{\todo}[1]{(\textbf{{#1}})\@latex@warning{Warning: todo used}}
\newcommand{\pgram}[0]{periodogram\xspace}

\setabbreviationstyle[acronym]{long-short}
\newacronym{ls}{L-S}{Lomb-Scargle}
\newcommand{\ls}[0]{\gls{ls}\xspace}
\newacronym{lsp}{LSP}{Lomb-Scargle Periodogram}
\newcommand{\lsp}[0]{\gls{lsp}\xspace}
\newacronym{ztf}{ZTF}{Zwicky Transient Facility}
\newcommand{\ztf}[0]{\gls{ztf}\xspace}
\newacronym{lsst}{SSPDB}{LSST Solar System Products Data Base}
\newcommand{\lsst}[0]{\gls{lsst}\xspace}
\newcommand{\adv}[0]{\textbf{Advantage}:\xspace}
\newcommand{\dadv}[0]{\textbf{Disadvantage}:\xspace}

\newacronym{vp}{VP}{VanderPlas}
\newcommand{\vp}[0]{\gls{vp}\xspace}
\newacronym{mc}{MC}{Monte Carlo}
\newcommand{\mc}[0]{\gls{mc}\xspace}

\newcommand{\tf}{$\SI{24}{\hour}$\xspace}

\biboptions{authoryear}
\newcolumntype{H}{>{\setbox0=\hbox\bgroup}c<{\egroup}@{}}
\usepackage[colorlinks=true]{hyperref} 

\begin{document}

\title{Removing Aliases in Time-Series Photometry}

\author[1,2]{Daniel Kramer\corref{cor1}}
\ead{drk98@nau.edu}
\author[1,2]{Michael Gowanlock}
\author[2,1]{David Trilling}
\author[2,4]{Andrew McNeill}
\author[3]{Nicolas Erasmus}

\cortext[cor1]{Corresponding author}
\address[1]{School of Informatics, Computing, and Cyber Systems, Northern Arizona University, Flagstaff, AZ 86011, USA}
\address[2]{Department of Astronomy and Planetary Science, Northern Arizona University, Flagstaff, AZ 86011, USA}
\address[4]{Department of Physics, Lehigh University, 16 Memorial Drive East, Bethlehem, PA 18015, USA}
\address[3]{South African Astronomical Observatory, Cape Town, 7925, South Africa}

\begin{abstract}
    Ground-based, all-sky astronomical surveys are imposed with an inevitable day-night cadence that can introduce aliases in period-finding methods.
    We examined four different methods --- three from the literature and a new one that we developed ---  that remove aliases to improve the accuracy of period-finding algorithms.
    We investigate the effectiveness of these methods in decreasing the fraction of aliased period solutions by applying them to the \ztf and the \lsst asteroid datasets.
    We find that the
    VanderPlas method had the worst accuracy for each survey. The mask and  our newly proposed window method yields the highest accuracy when averaged across both datasets. However, the Monte Carlo method had the highest accuracy for the \ztf dataset, while for \lsst, it had lower accuracy than the baseline where none of these methods are applied.
    Where possible, detailed de-aliasing studies should be carried out for every survey with a unique cadence.
\end{abstract}


\maketitle

\glsresetall
\glsunset{ztf}
\glsunset{lsst}
\glsunset{ls}

\section{Introduction}\label{sec:intro}

\sloppy
Determining periodic behavior among astrophysical sources is useful for describing their physical properties. For example, the internal strength of an asteroid can be determined using, among other observed properties, its rotation period \citep{mcneill2018} and light curves have been used to categorize different stellar types \citep{2022MNRAS.tmp.1449B}.

\fussy
Ground-based telescopic surveys that produce sparse data inevitably have
signals in the object's \pgram --- typically at or near $\SI{12}{\hour}, \SI{24}{\hour}, \SI{48}{\hour},$ and $ \SI{96}{\hour}$ --- related to the day-night cycle of the Earth.
These cadences cause aliasing, an effect where there are peaks in a \pgram that are not at the real period of an observed object.

If the output of a period-finding method and its corresponding light curve is visually examined, an astronomer can potentially make a judgment if a derived period is an alias or not. 
With large-scale surveys, like the Legacy Survey of Space and Time (LSST), too many objects will be observed for humans to manually confirm each derived period. This requires automating both deriving the periods and determining if the period is correct. 

The problem of period-finding at scale will become more acute when LSST is producing data, as
more than 100~million periodic sources are expected in the LSST catalog
\citep{lsstsciencebook}. If, for example, $20\%$ of period solutions are aliases, then 20~million sources would have incorrectly derived periods. The incorrect periods would either (1) be naively included in a catalog, (2) have solutions at common alias periods discarded, or (3) have to be examined with some verification algorithm. Any method that reduces the total number of alias results improves the three outcomes by (1) reducing the number of bad solutions, improving the overall accuracy, (2) increasing the overall number of correct solutions, or (3) decreasing the amount of computation to verify/score the periods. 

While this is an important problem for ongoing and upcoming surveys, there has been little progress on de-aliasing methods for sparse data. For instance, \citet{1992A&A...264..350C} has examined the de-aliasing properties of past methods, notably CLEAN \citep{1987AJ.....93..968R}. However, those methods only work on time series data with uniform observations during the night. Since modern ground-based surveys generate sparse data that is not uniform, these methods are unsuitable for these surveys. Recently, several papers have used some de-aliasing techniques using large scale survey data \citep{slow_rots, coughlin_ztf_2021, heinze_first_2018}, but none tested the improvement their method had over no de-aliasing.

In this paper, we analyze the performance of four approaches to de-aliasing, with performance and period-finding accuracy in mind.

The paper is organized as follows: Section~\ref{sec:lsp} discusses the Lomb-Scargle periodogram, Section~\ref{sec:datasets} gives an overview of the datasets used, Section~\ref{sec:methods} gives an overview of the four methods that were tested, Section~\ref{sec:results} discusses the results of the four methods, Section~\ref{sec:discussion} provides our discussion of the results of the paper, and Section~\ref{sec:conclusion} presents our conclusions and some future work on this and related problems.

\section{Period Finding Algorithms}\label{sec:lsp}

For testing the accuracy of the different de-aliasing methods, a period finding algorithm is needed. In this section, we will discuss four different methods, Lomb-Scargle (L-S, \citealt{1976Ap&SS..39..447L, 1982ApJ...263..835S}), SuperSmoother \citep{friedman_variable_1984}, Conditional Entropy \citep{graham_using_2013}, and Bootstrap $\chi^2$ \citep{2022FrASS...909771D}, and why \ls is used as the method for this analysis. 

\subsection{Lomb-Scargle}
\ls is a period-finding method first developed by \citet{1976Ap&SS..39..447L} and later improved by \citet{1982ApJ...263..835S}. It is one of the most popular period-finding algorithms used in astronomy. The general approach is to calculate the periodogram power, essentially a measure of the goodness of the solution, through Equation~\ref{equ:ls} as follows. 

\begin{align*}
\begin{split}
    P(f) &= \dfrac{1}{2}\left\{\dfrac{\Big(\sum_n g_n\cos(2\pi f[t_n-\tau])\Big)^2}{\sum_n\cos^2(2\pi f[t_n-\tau])}\right. \\&\qquad+\left. \dfrac{\Big(\sum_n g_n\sin(2\pi f[t_n-\tau])\Big)^2}{\sum_n\sin^2(2\pi f[t_n-\tau])}\right\} 
\end{split}\addtocounter{equation}{1}\tag{\theequation}\label{equ:ls}\\
    \tau &= \dfrac{1}{4\pi f}\tan^{-1}\left(\dfrac{\sum_n\sin(4\pi ft_n)}{\sum_n\cos(4\pi ft_n)}\right)\\
\end{align*}

Here, $P(f)$ is the power for an angular frequency $f$, $g_n$ is the observed telescopic magnitude of observation $n$ and $t_n$ is the time of observation $n$. Higher powers indicate a greater likelihood that $f$ is the angular frequency of the observed object. Note that the \lsp has a time complexity of $O(nm)$, where $n$ is the number of observations used and $m$ is the number of frequencies examined. 

\begin{figure}[!t]
    \centering
    \begin{subfigure}[b]{.5\textwidth}
    \includegraphics[width=\textwidth]{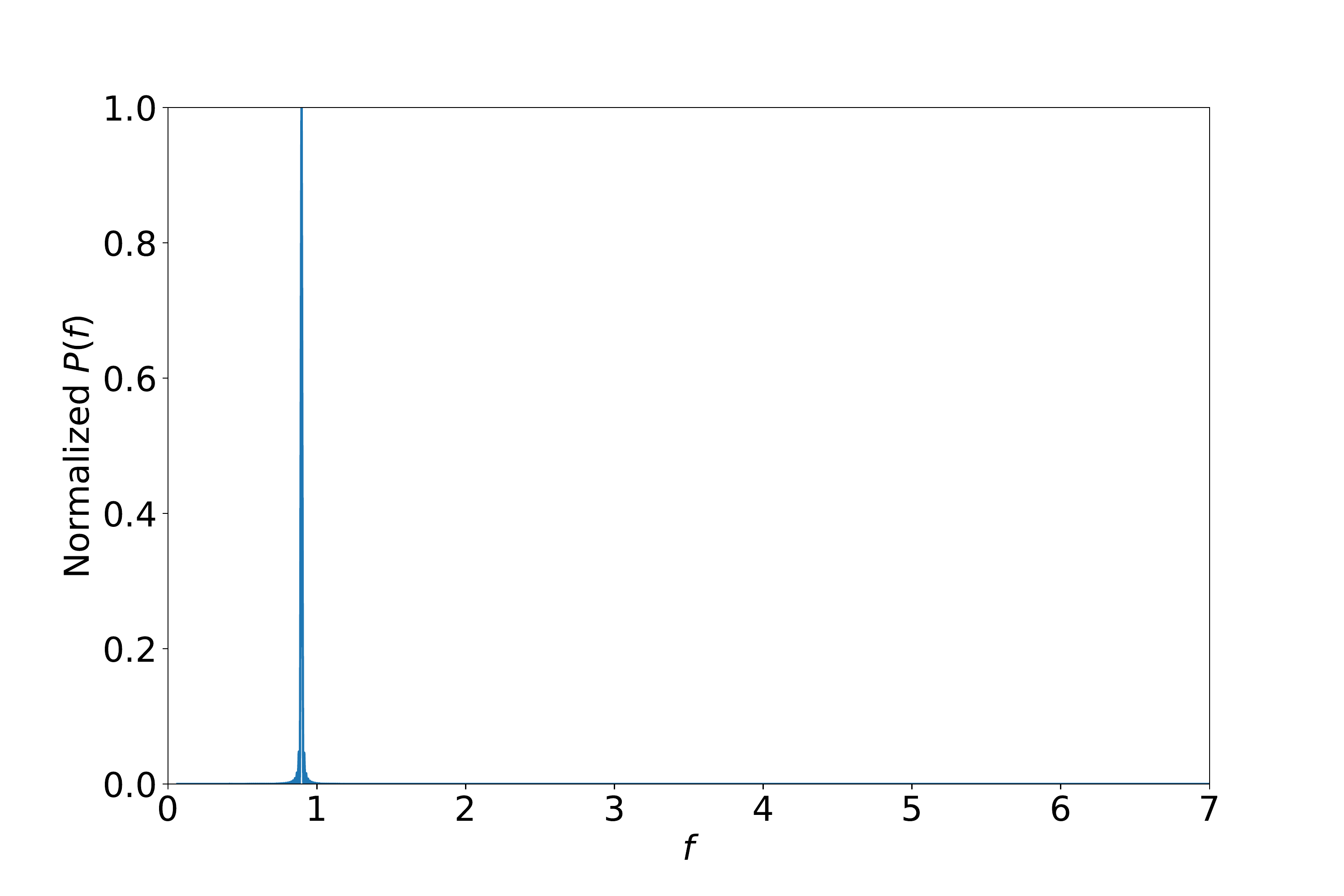}
    \caption{Uniformly sampled sine wave.}
    \end{subfigure}
    \begin{subfigure}[b]{.5\textwidth}
    \includegraphics[width=\textwidth]{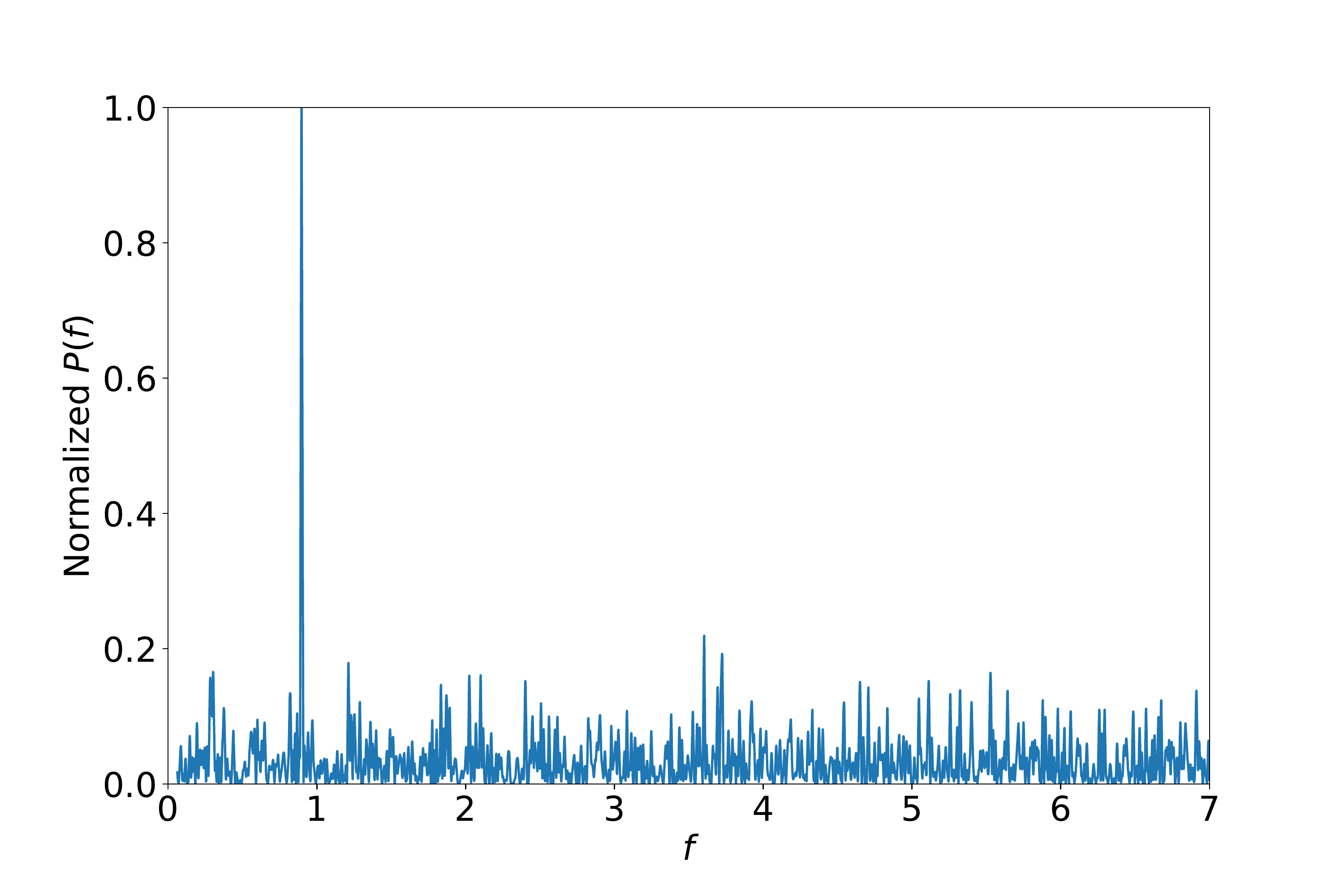}
    \caption{Randomly sampled sine wave.}
    \end{subfigure}
    \caption{The \pgram{s} for a uniformly sampled sine wave and a random sample of it. The x-axis is the angular frequency and the y-axis is the normalized power. The sine wave was given an angular frequency of $2\pi/7$ (period of $7$) and an amplitude of $0.3$. The pure sine wave used $\num{1000}$ points and the sampled sine wave used $0.5\%$ of those points, for about $50$ total points used.}
    \label{fig:pgram_comp}
\end{figure}

A \pgram is typically computed by calculating powers for a range of frequencies (or periods) in $[f_{\text{min}}, f_{\text{max}}]$ that are sampled over a uniform frequency space.
Ideally, there would only be a peak in the \pgram at the angular frequency $f$ that corresponds to the object's physical rotation state, with all other values of $P(f)$ having a value of zero. In reality, because of uncertainties in the measurements, a non-uniform cadence, and aliasing, the \pgram is often noisy with multiple peaks, so determining the correct peak is not always straightforward. Figure~\ref{fig:pgram_comp} shows how a \pgram generated with a uniformly sampled sine wave with a small $\Delta t$ compares to a randomly sampled sine wave's \pgram. The randomly sampled sine wave's \pgram is noisy around $P(f) = 0$ while the uniformly sampled sine wave only has a peak at the sine curve's frequency.

\subsection{Other Algorithms}
Other period finding algorithms exist, like SuperSmoother \citep{friedman_variable_1984}, Conditional Entropy \citep{graham_using_2013}, Bootstrap $\chi^2$ \citep{2022FrASS...909771D}, among others. With all of these algorithms, there is a trade-off between accuracy, speed, space, and light-curve shape flexibility. We summarize the three algorithms above as follows:
\begin{itemize}
    \item \emph{SuperSmoother} is especially useful for periodic signals that are not sinusoidal \citep{2011ApJ...731...17B, 2005AAS...20712216H}, has a time complexity and space complexity of $O(nm)$, and it is susceptible to aliasing \citep{gowanlock_gpu-enabled_2022}.

    \item \emph{Conditional Entropy} also has time complexity of $O(nm)$, is less accurate than \ls for fast periods, and is susceptible to aliasing \citep{coughlin_ztf_2021}.

    \item \emph{Bootstrap $\chi^2$} is the slowest, with a time complexity of $O(snm)$, where $s$ is the number of bootstrap samples, but it is the most accurate, with aliasing having the smallest effect on this method compared to the others described above \citep{2022FrASS...909771D}.
\end{itemize}

\ls has the same time complexity as SuperSmoother and Conditional Entropy, all three being smaller than Bootstrap $\chi^2$. As SuperSmoother has a space complexity of $O(nm)$ while \ls and Conditional Entropy have a space complexity of $O(n+m)\approx O(m)$, SuperSmoother was not used. Finally, since L-S and Conditional Entropy are similar, and \ls is the the standard algorithm used for deriving the periods in astronomy, we elect to use \ls in this paper. However, any period finding algorithm that produces a \pgram is capable of implementing the methods described in Section~\ref{sec:methods}.

\section{Data Used}\label{sec:datasets}
The upcoming LSST --- to be carried out with the Vera C. Rubin Observatory that is presently under construction in Chile --- will revolutionize many fields of astronomy \citep{2019ApJ...873..111I}. LSST will generate sparse photometry on the around 500--1000~measurements over ten~years for some 40 billion astronomical sources. The vast majority of these will be sidereal objects (stationary on the sky); around 5 million of them will be moving objects. Many of these LSST-observed sources will be variable, with regular periods, so it is of interest to develop and implement accurate period-finding algorithms that can operate at the vast LSST scale. 

The ongoing all-sky survey being carried out by the Zwicky Transient Facility (ZTF; \citealt{Bellm_2018}) acts as a kind of LSST precursor. \ztf is carrying out a public survey that is very LSST-like in terms of cadence, data type, and data accessibility, but at something like one-tenth the LSST scale.

The work presented here has its origin in Solar System science and asteroid period finding but is relevant for period-searching for any kind of astrophysical source in either \ztf or LSST data.
For \ztf, we used asteroid data from SNAPShot1 \citep{Trilling_2023}, which used \ztf observations from 2018--07--19 to 2020--05--19, using only numbered asteroids with more than 50~observations  with Real-Bogus scores $\geq0.55$ \citep{10.1093/mnras/stz2357}; the Real-Bogus cut eliminated about $9\%$ of all observations.

There is no actual LSST data yet as science operations will commence in 2024, so we used the LSST Solar System Products Data Base (SSPDB; \citealt{Juric2021Being}), a complete simulation of asteroid observations over the 10~year nominal lifetime of LSST, as our LSST testbed. However, the objects in this synthetic database are all assumed to be spherical and thus, unlike almost all real asteroids, do not show shape-induced periodic variability in the lightcurves. We therefore assigned lightcurve amplitudes and rotation periods to each \lsst object, as described in Appendix~\ref{sec:lsst_gen}.

\section{Methods}\label{sec:methods}

In this section, 
we present our de-aliasing analysis on the \ztf and \lsst asteroid data sets. We use three methods from the literature --- masking; \mc; and \vp --- and present our new approach, the window method. Broadly speaking, the \mc method tries to remove aliasing through subsampling (with multiple trials), whereas the other methods attempt to remove signals at the expected alias periods.

One method that will not be tested is the False Alarm Probability \citep{2008MNRAS.385.1279B}. False Alarm Probability is sometimes incorrectly used as a proxy for the goodness of a \ls solution. However, False Alarm Probability is a way of calculating the probability $p=(\text{power}|\text{noise})$: the probability that a given \pgram power is part of the noise of the \pgram \citep{2008MNRAS.385.1279B,vp_ls}. If there is an alias peak with a larger $P(f)$ than the real peak, then it would have a lower False Alarm Probability. Because False Alarm Probability is not testing for the authenticity of a signal related to the physical period in the system being monitored, it is not a useful method for determining if a peak is a real period or an alias.

\subsection{Masking}
\label{subsec:mask}
Masking is the most straightforward approach presented in this paper: solutions near the known alias solutions are simply rejected.
The alias periods or a small range around the alias can be removed (masked out) from the \pgram so the real period's peak would therefore have the largest remaining $P(f)$.
Usually, there are multiple aliases, so several masks are needed to remove them.

This method was used in \citet{slow_rots}, which presented asteroid photometry from Asteroid Terrestrial-impact Last Alert System and \ztf where they masked out periods of \{8, 12, 16, 24, 48\}~hours. 
The remaining periodogram peaks with the largest power were then found to be those representing super-slow rotation periods. \citet{slow_rots} showed that the masking method is a viable way of removing alias period solutions so we incorporate this method into our analysis. This method was also used in \citet{coughlin_ztf_2021}, which used a similar method to derive the mask ranges as described below. The \citet{coughlin_ztf_2021} mask ranges, in rotation period space, are 
[(0.5, 0.5),
 (0.51, 0.51),
 (0.52, 0.52),
 (5.93, 6.08),
 (7.87, 8.14),
 (11.71, 12.31),
 (22.86, 25.26),
 (46.15, 50.0),
 (600.0, 800.0)]~hours.

\begin{table}[!t]
\centering
\begin{tabular}{|l|p{.375\textwidth}|}\hline
    Dataset & Mask Ranges\\\hline
    \ztf & (4.789, 4.814),
 (5.989, 6.014),
 (7.972, 8.014),
 (11.947, 12.039),
 (23.764, 24.164) \\\hline
    \lsst & (4.789, 4.839),
 (5.939, 6.039),
 (7.939, 8.039),
 (11.889, 12.039),
 (23.639, 24.239) \\\hline
\end{tabular}

\caption{\label{tab:mask_ranges}The period ranges (exclusive) for each of the datasets to be excluded from the (light-curve) period space. Note that these ranges are light curve periods, not rotation periods, where an object's light curve period is half its rotation period.}
\end{table}

\begin{figure}[!t]
    \centering
    \includegraphics[width=.5\textwidth]{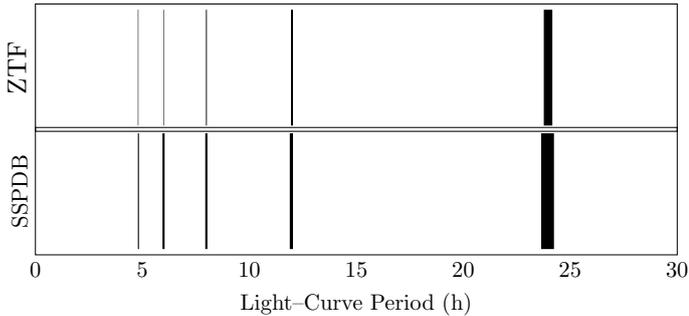}
    \caption{Visualization of the mask ranges from Table~\ref{tab:mask_ranges}. The black bars indicate the masked regions where all solutions are rejected. \ztf and \lsst have similar, but not exactly the same, ranges.
    }
    \label{fig:mask_ranges}
\end{figure}

\begin{figure}[!t]
    \centering
    \includegraphics[width=.5\textwidth]{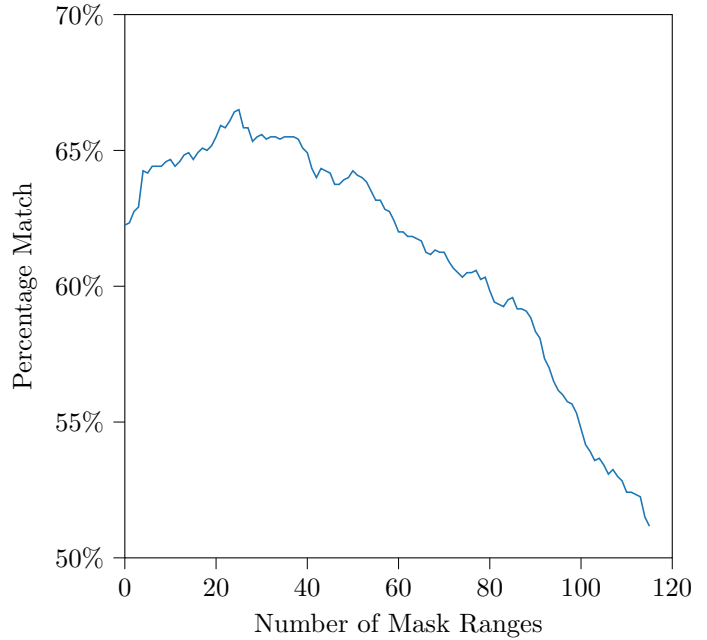}
    \caption{The match percentage for each new mask range. The x-axis is the number of $\SI{0.05}{\hour}$ size mask ranges. The y-axis is the match percentage.}
    \label{fig:mask_ranges_deriv}
\end{figure}

Table~\ref{tab:mask_ranges} and Figure~\ref{fig:mask_ranges} show the period ranges that are masked out for each dataset. The mask ranges were derived through the following steps:

\begin{enumerate}
    \item Generate a histogram of the derived periods.\label{list:mr_steps_1}
    \item Create a new mask range by selecting the bin with the highest number of objects. For example, if the bin with the most objects is $[\SI{23.9}{\hour},\SI{24.1}{\hour}]$, then it would be added as a mask range.
    \item Re-derive the periods with this new mask range. \label{list:mr_steps_3}
    \item Repeat steps~\ref{list:mr_steps_1} through~\ref{list:mr_steps_3} consecutively, adding new masks. Then select the mask ranges that provide the maximum match percentage to a database of accurate periods for the objects. For example, Figure~\ref{fig:mask_ranges_deriv} shows percent match against the Light Curve Database (LCDB; \citealt{https://doi.org/10.26033/j3xc-3359}) as a function of the number of masks for \ztf.
\end{enumerate}

The ranges for both \ztf and \lsst are similar (Table~\ref{tab:mask_ranges} and Figure~\ref{fig:mask_ranges}), but not all surveys will have these same ranges because they will have different cadences and different observational errors. Despite this, we expect that all ground-based surveys will have the \tf alias.

\adv The primary advantage of this approach is that it is the fastest to compute. The time complexity for the method is $O(nm)$, where $n$ is the number of data points used and $m$ is the number of frequencies examined. If the masks are pre-computed, the run time is slightly faster compared to a normal \lsp because, with the same frequency range and $\Delta f$, fewer frequencies ($m$) would have to be examined as the frequencies in the mask ranges would be excluded.

\dadv This method has the disadvantage that true periods that are within one of the masked ranges will never be identified.
 The LCDB contains a large number of curated asteroid rotation periods and only about $1\%-2\%$ of objects in the LCDB have rotation periods that are in the masked ranges of Table~\ref{tab:mask_ranges}, so only a low number of objects would be impacted. 

\subsection{Monte Carlo (MC)}
Since aliases are generated by the observing cadence intervals, randomly subsampling the observations, across many trials, might suppress the signal from the aliases.
An overall \pgram for each asteroid can then be found either by finding the most common peak power from each trial's \pgram, or by  summing over all the \pgram{}s. Some attention must be paid to ensure that the time baseline of the subsample ($T_{\text{sub}}$) is greater than $P_{\text{max}}$, the period corresponding to the minimum frequency examined, otherwise, \ls will derive the period of a partial light-curve, causing an incorrect period to be derived. 
For large scale surveys, like \ztf and LSST, that produce sparse photometry over years, this is not an issue as $T_{\text{sub}}$ will always be greater than $P_{\text{max}}$ for any reasonable $P_{\text{max}}$.
   
For the calculations in this paper, we used 100 selection samples/trials ($i$) per object with 50 random observations ($n$). \lsst objects had to have at least 200 observations while \ztf objects only had to have 90 observations so there would be enough objects to get a large statistical sample. These parameters were found via an ad hoc process of changing the minimum number of observations required, the number of observations per subsample, and the number of \mc trials to find a high match percentage. 

Finally, we sum the un-normalized \pgram{s} over all trials to yield the overall \pgram. The highest peak from this \pgram is used as the derived period. 

\adv This method has the advantage that, compared to the other methods, it is the most oblivious to aliases like those in Table~\ref{tab:mask_ranges}. Unlike the masking method, the \mc method could derive a correct rotation period that happens to be an alias. 

\dadv This method also requires more observations than the other methods discussed because in order to produce a credible \pgram, the number of observations in each subsample needs to be sufficiently large while also being significantly smaller than the total number of data points. This is needed so that pairs of observations that are, for example, \tf apart are excluded from the same subsample. 

This method is also the slowest, with a time complexity of $O(nmi)$, where $n$ is the number of data points used for each iteration, $m$ is the number of frequencies checked, and $i$ is the number of selection samples.

\subsection{Window} 

The window function derives aliases caused by the observational cadence. It takes the temporal data from an object's observations to produce a \pgram-like output where a frequency having a high power corresponds to high aliasing \citep{vp_ls}.
\begin{align}
\label{equ:window}
    \mathcal{P}_w(f;\{t_n\})=\left|\sum_{n=1}^N e^{-2\pi ift_n}\right|^2
\end{align}
Equation~\ref{equ:window} shows the window function, but \citet{vp_ls} showed that a \lsp can be used as an approximation of the window function if in Equation~\ref{equ:ls}, $g_n=1$; this simulates a completely spherical, homogeneous object. 
Therefore, any signal present in the window \pgram would be aliasing caused by the underlying cadence. 

\begin{figure}
    \centering
    \includegraphics[width=.5\textwidth]{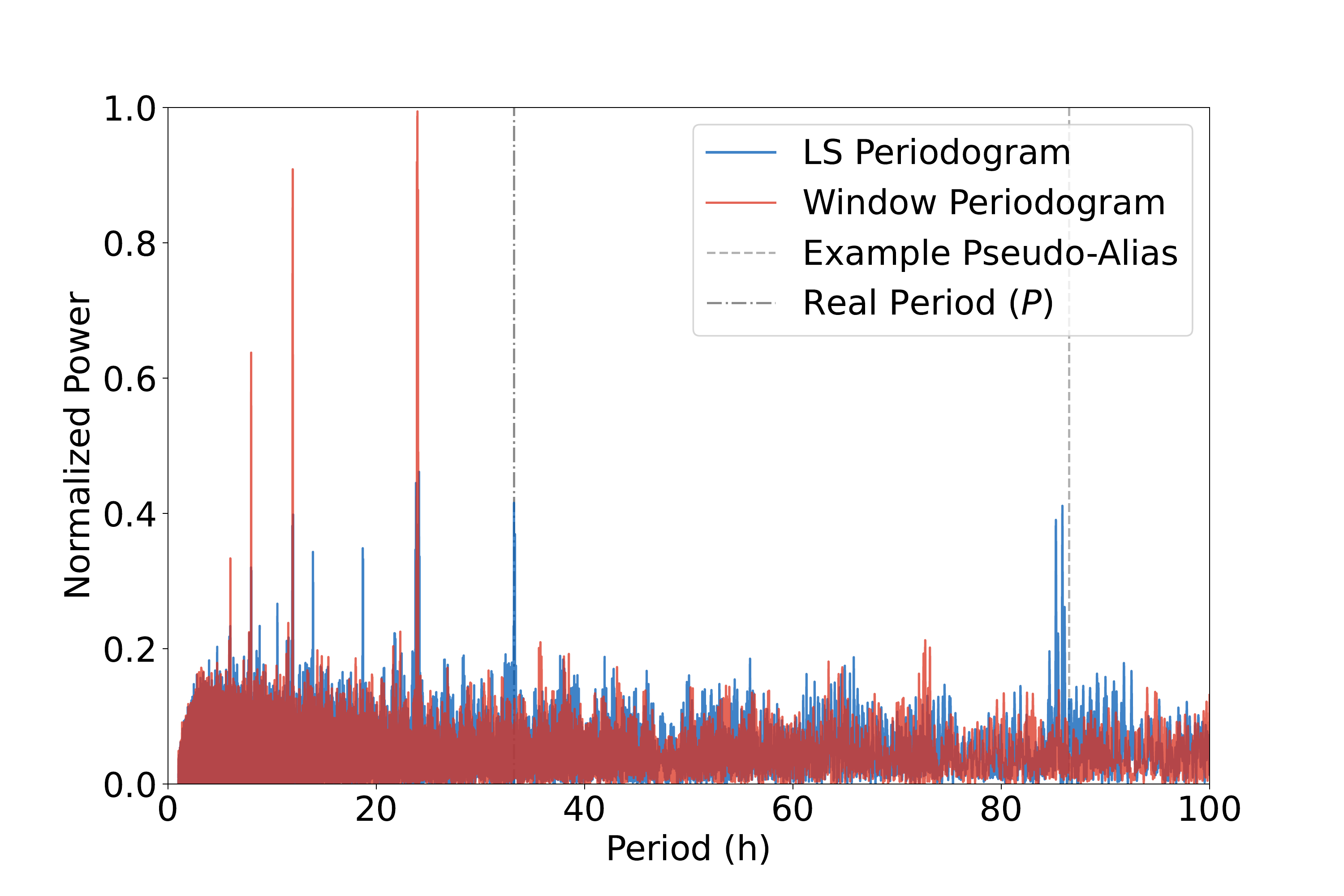}
    \caption{The \lsp and window \pgram for an object from \lsst. The \lsp is in red and the window \pgram is in blue. The given period for the object is marked by the dash-dot-dash line and its peak is only present in the \lsp. The dashed line marks an example pseudo-alias which is also only present in the \lsp. The peaks in the window \pgram are the aliases which are also all present in the \lsp.
    }
    \label{fig:window_example}
\end{figure}

Figure~\ref{fig:window_example} provides an example of how an object's \lsp and window \pgram relate. 
There are three key observations from Figure~\ref{fig:window_example}: (1) there are peaks in the \lsp that do not correspond to the real period, (2) all of the significantly strong peaks that are in the window \pgram also appear in the \lsp, and (3) there exist peaks in the \lsp that are not the real period or in the window \pgram; these are pseudo-aliases. The example pseudo-alias marked in Figure~\ref{fig:window_example} is $\left((1/P)-(1/\SI{24}{\hour})\right)^{-1}$ where $P$ is the real period for the object \citep{2022FrASS...909771D}.

Previous algorithms that use the window function, like deconvolution and CLEAN, do not work for removing aliases \citep{vp_ls} because they assume that the strongest peak in the \pgram is the peak that corresponds to the real period. Here we present a new way to use the window function to remove aliases, where the pseudocode can be found in Algorithm~\ref{algo:window}. This differs from deconvolution and CLEAN because they rely on deconvolution for their alias removal while Algorithm~\ref{algo:window} does not.

\begin{algorithm}
    \caption{Our method of using the peaks in the window \pgram to check if a peak in the \lsp is an alias.}
    \label{algo:window}
    \begin{algorithmic}[1]
    \Procedure{windowMethod}{time, mag, freqs}
    \State pgramLS $\gets$ LS(time, mag, freqs) \label{algo:window:ls}
    \State pgramWindow $\gets$ Window(time, freqs) \label{algo:window:window}
    \State LSPeaks $\gets$ findPeaks(pgramLS) \label{algo:window:peakLs}
    \State WindowPeaks $\gets$ findPeaks(pgramWindow) \label{algo:window:peakWindow}
    \State sort(LSPeaks) \label{algo:window:sort}
    
    \State NPeaks $\gets$ length(LSPeaks)
    \State index $\gets 0$
    
    \While{index $<$ NPeaks \textbf{and} LSPeaks[index] $\not\in$ WindowPeaks} \label{algo:window:while}
        \State index $\gets$ index$+1$
    \EndWhile{}
    \If{index $\geq$ length(LSPeaks)} \label{algo:window:if}
        \State correctPeak $\gets$ LSPeaks[0]
    \Else
        \State correctPeak $\gets$ LSPeak[index]
    \EndIf
    \State \Return correctPeak
    \EndProcedure
    
    \end{algorithmic}
\end{algorithm}

Algorithm~\ref{algo:window} takes an object's time of observation $T=\{t_1, t_2, \ldots, t_n\}$ (time), observed magnitudes $G=\{g_1, g_2, \ldots, g_n\}$ (mag) where $n$ is the number of observations, and the frequency grid on $[f_{\text{min}}, f_{\text{max}}]$ (freqs) as arguments. The \texttt{LS} and \texttt{Window} functions on lines~\ref{algo:window:ls} and \ref{algo:window:window} calculate the \ls and window \pgram{s} respectively. The \texttt{findPeaks} methods on lines~\ref{algo:window:peakLs} and \ref{algo:window:peakWindow} find the peaks in those \pgram{s} and returns the peak frequencies and their powers. Line~\ref{algo:window:sort} sorts the \lsp peaks by their power in descending order.
Line~\ref{algo:window:while} loops until the current \lsp peak is not contained in the window peaks. 
If the \lsp peak is in the set of window peaks, the index of the current peak is incremented. Line~\ref{algo:window:if} tests if all of the detected \lsp peaks have been compared against the set of window peaks. If they have, then the peak with the highest power is used as the correct peak (although this is probably an alias), otherwise, the peak at the current index is used as the correct peak. The correct peak is then returned. A python implementation is available as a \href{https://gist.github.com/drk98/6b15633e8fc43e8daf6b628548006376}{GitHub gist here}\footnote{\url{https://gist.github.com/drk98/6b15633e8fc43e8daf6b628548006376}}.
   
\adv This method has a distinct advantage over the masking method. The window function exposes aliases on a per-object basis, whereas the masking method uses a single set of masks for all objects in a catalog. Consequently, the window method may enable finding correct periods that are typical aliases (e.g., those defined in Section~\ref{sec:intro}), whereas the masking method excludes all of these periods.

\dadv Using Algorithm~\ref{algo:window}, this method has the same time complexity for \ls as the other methods, which is $O(nm)$. It also has an additional time complexity for the peak finding algorithm, $\Psi$, which we describe below. However, since this method requires two \lsp to be calculated, it is at least twice as many operations as the \lsp. The implementation for the window \pgram peak finding function considered any power greater than $5\sigma$ from $0$, where $\sigma$ is the standard deviation of all the powers in the window \pgram, a peak. The \lsp peaks were found using the \texttt{argrelmax} function in SciPy/cuSignal \citep{2020SciPy-NMeth,RAPIDS}, which has a time complexity $\Psi$ of $O(om)$ where $m$ is the number of frequencies examined and $o$ is the \texttt{order} parameter. The \texttt{order} parameter for \texttt{argrelmax} and the minimum peak height, relative to the strongest peak, is located in Table~\ref{tab:window_params}.

\begin{table}[!t]
    \centering
    \begin{tabular}{ccc}
         & \begin{tabular}{@{}c@{}}\ls\\ Min. Peak Height\end{tabular}& \begin{tabular}{@{}c@{}}\ls\\ Peak Order\end{tabular}  \\\hline
        \ztf & $1/7$ & $\num{10000}$ \\

        \lsst & $1/10$&$\num{3000}$
    \end{tabular}
    \caption{The parameters for the window/VanderPlas methods. ``\ls Min. Peak Height'' is the minimum power of a peak, relative to the highest power, to still be considered a peak. For example, for LSST, if the highest peak's power was $0.8$, then the peak power cutoff would be $0.08$. ``\ls Peak Order'' is the ``order'' argument to the \texttt{argrelmax} function in scipy/cuSignal.}
    \label{tab:window_params}
\end{table}

\subsection{VanderPlas (VP) Method}
\citet{vp_ls} describes a method for removing aliases that also utilizes the window function. The difference between this method and the window method is that it compares more peaks between the \lsp and window \pgram and in particular it considers pseduo-aliases.

The \vp method steps are as follows, where $f_{\text{peak}}=\text{max}(P(f))$:
\begin{enumerate}
    \item Check if there are any peaks in the \lsp at $f_{\text{peak}}/m, $ where $m\in\{2,3\}$ which checks if the found peak is an integer multiple of the real peak. 
    \item Check for peaks at $f_{\text{peak}}\pm n\delta f,$ where $n\in\{1,2\}$ and where $\delta f$ is the frequency having the highest window \pgram power.
    \item Manually check the highest peaks and fit a model to find the best one. We ignore this step as this requires human intervention, which is not feasible for large-scale survey data.

\end{enumerate}

This method has the same time complexity as the window method at $O(nm)+\Psi$, where $\Psi$, when using \texttt{argrelmax}, is $O(om)$ and it shares its \textbf{advantages} and \textbf{disadvantages} since the methods are similar.
The implementation also uses the same parameters as the Window method, located in Table~\ref{tab:window_params}.

\section{Results}\label{sec:results}

All \ls and window functions used the implementations from SciPy/cuSignal, using $\num{5e6}$ frequencies on a uniform period grid on $[\SI{1}{\hour}, \SI{150}{\hour}]$. 

As a baseline, Figure~\ref{fig:no_meth_dist} shows the period distribution when no dealiasing is applied and  Figures~\ref{fig:mask_dist}--\ref{fig:vdp_dist} shows the period distributions for the four methods presented above. 

Each of the distributions has spikes at aliases, meaning that none of the approaches are completely successful at removing all aliases because we assume that the true period distributions for our population are continuous .

Our next step is to compare the derived results with known values. For \ztf period solutions, the known values are provided by the LCDB (Section~\ref{subsec:mask}), which contains few alias solutions because of human curating and/or observational cadence that do not have aliasing at the derived periods. There are $\num{2544}$ objects that are in both the LCDB and \ztf; all of them were used for the calculations. 
For the \lsst, since we assigned rotation periods for every object, verifying the correct periods in aliased and de-aliased solutions for this case is straightforward. 

The percentage match between the derived period and the real period is shown in Table~\ref{tab:lsst_match}. Objects were considered to have matching periods if their period, half their period, or double their period was within 10\% of the LCDB (for \ztf) or assigned period (for \lsst). The baseline method for both surveys had a match rate of over $50\%$, so most objects have their correct period derived. Once the methods were applied, the masking and window methods increased the match percentage over the baseline for both surveys while the \mc method only increased the match percentage for \ztf, and the \vp method decreased the match percentage for both surveys.

\begin{table}[!t]
\centering
\begin{tabular}{cHrrrr}
\hline
\multirow{2}{*}{Method} & \multicolumn{3}{c}{Percentage Match} & \multicolumn{2}{l}{Percentage Change}              \\
                        & ATLAS      & \ztf       & \lsst      & \multicolumn{1}{r}{\ztf} & \multicolumn{1}{r}{\lsst} \\ \hline
None/Baseline           & $24.0\%$  & $64.8\%$  & $57.9\%$  & ---                 & ---                \\ \hline
Mask                    & $29.3\%$  & $65.8\%$  & $74.5\%$  & $1.56\%$                 & $28.6\%$                 \\ \hline
\mc                     & $33.0\%$  & $70.1\%$  & $53.5\%$  & $8.08\%$                 & $-7.61\%$                 \\ \hline
Window                  & $30.2\%$  & $69.0\%$  & $71.5\%$  & $6.39\%$                 & $23.4\%$                  \\ \hline
\vp                     & $\%$       & $14.0\%$  & $8.91\%$   & $-78.4\%$               & $-84.6\%$                \\ \hline
\end{tabular}
\caption{The percentage match between the derived period and the real period for the three surveys and the percentage change from the baseline. The ``None/Baseline'' row is when none of the methods were used and the highest \lsp power was used as the derived period. \lsst used a sample of $\num{10000}$ objects.}
    \label{tab:lsst_match}
\end{table}

\section{Discussion}\label{sec:discussion}

We begin by examining \ztf and excluding the \vp method (which will be described later). The best match percentage using \ztf is with the \mc method; however, all of the methods are within $\approx6\%$ of each other, including the baseline. At a minimum, if de-aliasing is needed for a \ztf-like survey, the masking method should be implemented as it incurs no extra cost. If computational cost is not prohibitive, then the \mc should be implemented. In all cases (except where VP is implemented), our results suggest that around two-thirds of all reported solutions are likely to be correct.

For \lsst, 
the variation among non-VP approaches is greater than for \ztf, ranging from barely better than 50\% (MC) to almost 75\% accurate (masking).
Notably, 
the \mc method performs worse than the baseline. We hypothesize that the long baseline of the \lsst observations leads to the points in each subsample being too temporally distant\footnote{This is only an issue for the \mc method. Having a long baseline of observations is beneficial for the other methods.} for \ls to reliably derive the correct period. 

The \vp method was the worst performing of the methods as it seems to incorrectly select aliases/pseudo-aliases, causing the method to ``overcorrect''. 
Figure~\ref{fig:vdp_dist} shows that there are regions where no periods are detected 
and periods tend to be derived at longer periods, meaning that the correct periods get derived a small percentage of the time. With the \vp and window methods being similar, we hypothesize that the reason the \vp method is worse is that the pseudo-alias check causes an overcorrection of the periods. 

One may wonder whether the match percentages found here could be improved since all of the methods still have derived periods at or near aliases. It is possible that better parameters for the methods could be found with a more exhaustive search of the parameter space, like those in Table~\ref{tab:window_params}. However, such a search is impractical due to the large volume of data in the catalogs.

For the masking method, since it is static and therefore unable to react to changes in aliases, the masks might have to be re-derived after the survey starts if an inaccurate simulation was used or if the observational cadence of the survey changes during the survey.

One important conclusion is that the best de-aliasing approach is not the same for \ztf and \lsst, which implies that a study like this should be carried out for every large-scale survey. If this is impractical and a single uniform approach is preferred, we identify the masking process as the most effective, though this conclusion is based only on the two data sets considered here.

\section{Conclusions and Future Work}\label{sec:conclusion}

We used two sets of survey data, one real and one synthetic, 
to test four de-aliasing techniques for period solutions from all-sky surveys.

We find that the masking method provides the overall best results and should be chosen for any given survey. This method has a relatively low time complexity.
The masks for \ztf and \lsst, and therefore LSST, have been generated and are presented in Table~\ref{tab:mask_ranges}. The window method would also be a good choice to apply if the computational performance loss compared to the masking method is not important as it provides aliases for each object individually.
However, we note that results may vary from survey to survey, and the best approach is to carry out individualized studies, such as this one.

This paper leads to several lines of future investigation:
\begin{itemize}
    \item Improving the window method so it provides a higher match rate. 
    \item Use several of the methods together to see if that improves the match rate.
    \item Develop a GPU version of the \mc method in order to decrease its expensive computation time.
    \item Develop a method that removes the pseudo-aliases.
    \item Test how differences in a survey's simulated data and its real data change its mask ranges.
    \item Develop and test methods for determining a ``confidence'' in a \pgram result in order to better gauge if a derived period result is correct.
    
\end{itemize}

\section{Acknowledgements}

We acknowledge many useful conversations with Nat Butler and John Kececioglu and with Tom Matheson and the ANTARES team.
This work has significantly benefited from all of their expertise.

This work has been supported in part by the Arizona Board of Regents, Regents’ Innovation Fund.

ZTF is a public-private partnership, with equal support from the ZTF Partnership and from the U.S. National Science Foundation through the Mid-Scale Innovations Program (MSIP). The ZTF partnership is a consortium of the following universities and institutions (listed in descending longitude): TANGO Consortium of Taiwan; Weizmann Institute of Sciences, Israel; Oskar Klein Center, Stockholm University, Sweden; Deutsches Elektronen-Synchrotron \& Humboldt University, Germany; Ruhr University, Germany; Institut national de physique nucl\'eaire et de physique des particules, France; University of Warwick, UK; Trinity College, Dublin, Ireland; University of Maryland, College Park, USA; Northwestern University, Evanston, USA; University of Wisconsin, Milwaukee, USA; Lawrence Livermore National Laboratory, USA; IPAC, Caltech, USA; Caltech, USA.
We thank the anonymous reviewer for their insightful comments and helpful feedback on our manuscript.

\bibliography{refs}

\begin{appendices}

\section{LSST Generation}
\label{sec:lsst_gen}
\begin{figure}
    \centering
    \begin{subfigure}[b]{.5\textwidth}

        \includegraphics[width=\textwidth]{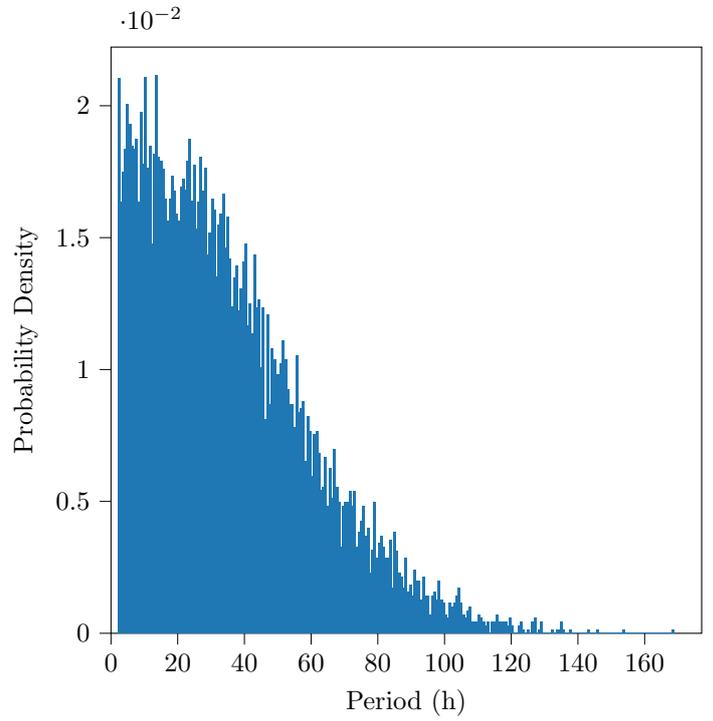}
        \caption{The period distribution used for \lsst.}
    \end{subfigure}
    \begin{subfigure}[b]{.5\textwidth}

        \includegraphics[width=\textwidth]{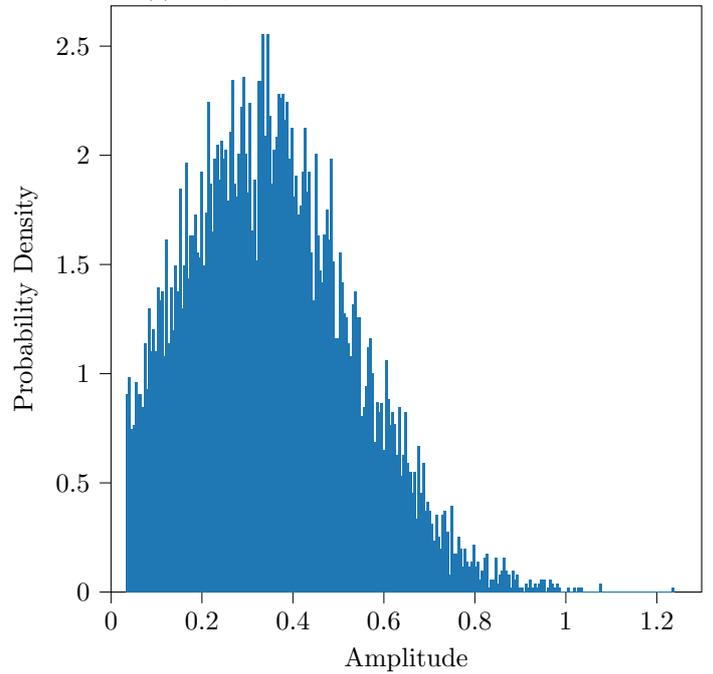}

        \caption{The amplitude distribution used for \lsst.}
    \end{subfigure}
    \caption{The period and amplitude distributions for \lsst. The x-axis is the period/amplitude and the y--axis is the probability density of each bin (the area under the histogram is $1$).}
    \label{fig:lsst_p/a_dist}
\end{figure}

For each object from the LSST Synthetic Moving Objects Database, the following cuts/steps were taken in generating a periodic signal in their data
\begin{enumerate}
    \item Import all objects with at least 30 observations in at least two filters.
    
    \item Convert the apparent magnitudes to absolute magnitudes using the ``filterg12" value as that filter's $G$. \lsst uses the Bowell HG system \citep{1989aste.conf..524B} not the HG12 system \citep{2010Icar..209..542M} even though the field is called ``filterg12". If that filter's ``g12" data was NaN, then the average of the rest of the filter ``g12" values were used. If all the filter's ``g12" values are NaN, then a value of $G=0.15$ was used for all filters.

    \item A period and amplitude were generated for each object and a sine wave with those properties was added to the objects derived absolute magnitudes. Both were generated using the parameters located in Table~\ref{tab:lsst_lc_params}  and the resulting distributions are shown in Figure~\ref{fig:lsst_p/a_dist}. The truncated normal distribution used SciPy's \texttt{truncnorm} function and the gamma distribution used SciPy's \texttt{gamma} function.
    \begin{table}[ht]
        \centering
        \begin{tabular}{l|ll|ll}
                            & \multicolumn{2}{l|}{Period}           & \multicolumn{2}{l}{Amplitude} \\ \hline
Distribution                & \multicolumn{2}{l|}{Truncated Normal} & \multicolumn{2}{l}{Gamma}     \\ \hline
\multirow{4}{*}{Parameters} & $a$                & $-0.1$              & $a$           & $0.48$         \\
                            & $b$                  & $50$               & $\frac{1}{\beta}$         & $0.18$           \\
                            & $\mu$                 & $6$                &               &               \\
                            & $\sigma$              & $40$               &               &              
\end{tabular}
        \caption{The distributions and parameters used for the period and amplitude generation. Plots for these distributions are available in Figure~\ref{fig:lsst_p/a_dist}. The distributions were approximated from the LCDB data (Section~\ref{subsec:mask}).}
        \label{tab:lsst_lc_params}
    \end{table}
    
    \item For each filter, the mean of the absolute magnitudes, $\mu_H$, was calculated. $H-\mu_H$ for each filter was then concatenated, resulting in a single band of data with $\mu_H=0$.
    
    \item The observation data along with the assigned period and amplitude values were stored. 
    
\end{enumerate}

\section{Derived Period Distributions}

\begin{figure}[H]
    \centering
    \begin{subfigure}[b]{.5\textwidth}

    \includegraphics[width=\textwidth]{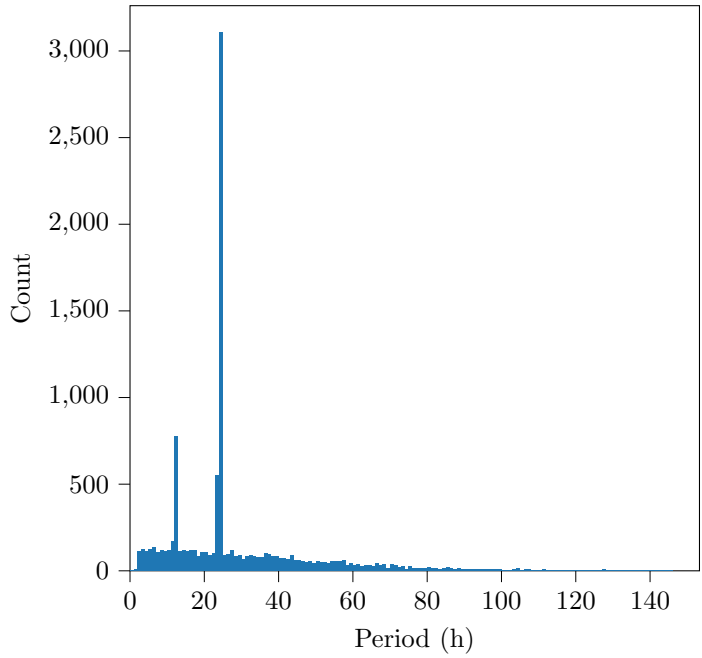}

    \caption{\lsst}
    \end{subfigure}
    \begin{subfigure}[b]{.5\textwidth}
  
    \includegraphics[width=\textwidth]{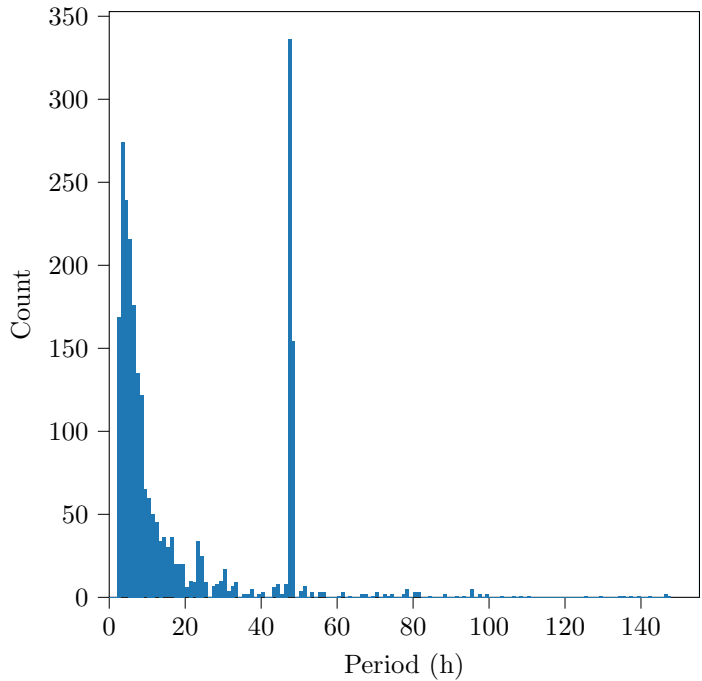}
    \caption{\ztf}
    \end{subfigure}
    \caption{The derived period distribution for the given survey using the no method (the base derived period distribution).}
    \label{fig:no_meth_dist}
\end{figure}

\begin{figure}[H]
    \centering
    \begin{subfigure}[b]{.5\textwidth}

    \includegraphics[width=\textwidth]{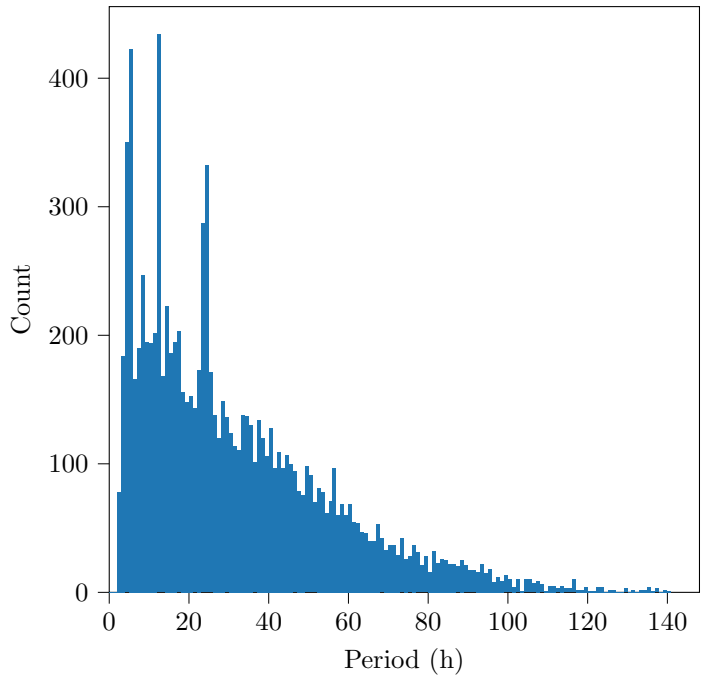}
    \caption{\lsst}
    \end{subfigure}
    \begin{subfigure}[b]{.5\textwidth}
  
    \includegraphics[width=\textwidth]{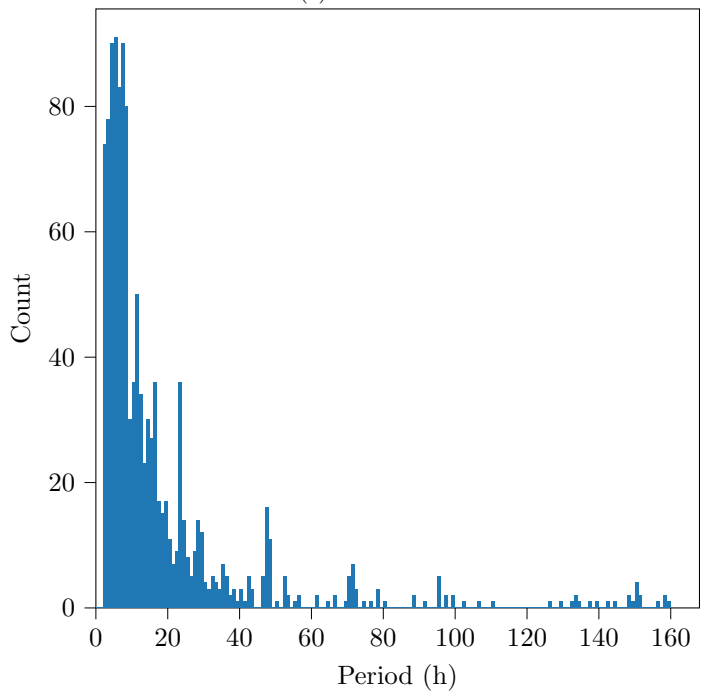}
    \caption{\ztf}
    \end{subfigure}
    \caption{The derived period distribution for the given survey using the mask method.}
    \label{fig:mask_dist}
\end{figure}

\begin{figure}[H]
    \centering
    \begin{subfigure}[b]{.5\textwidth}

    \includegraphics[width=\textwidth]{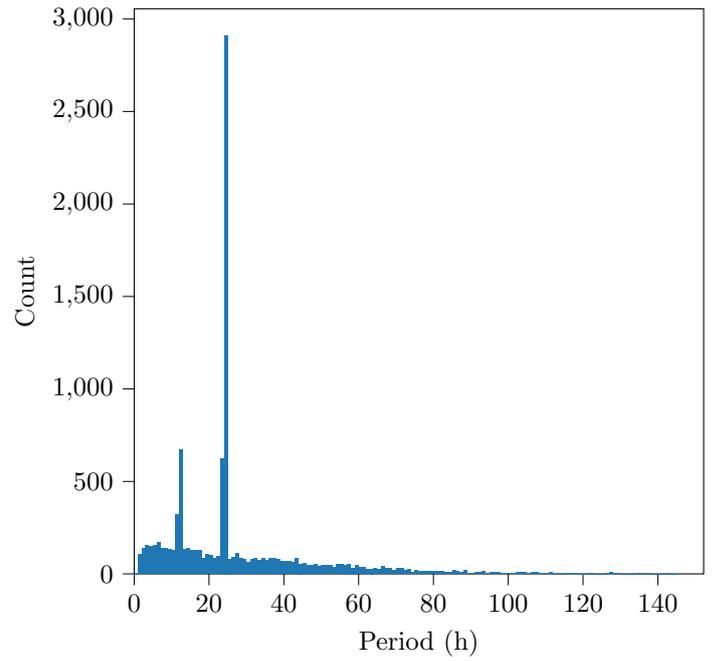}

    \caption{\lsst}
    \end{subfigure}
    \begin{subfigure}[b]{.5\textwidth}
  
    \includegraphics[width=\textwidth]{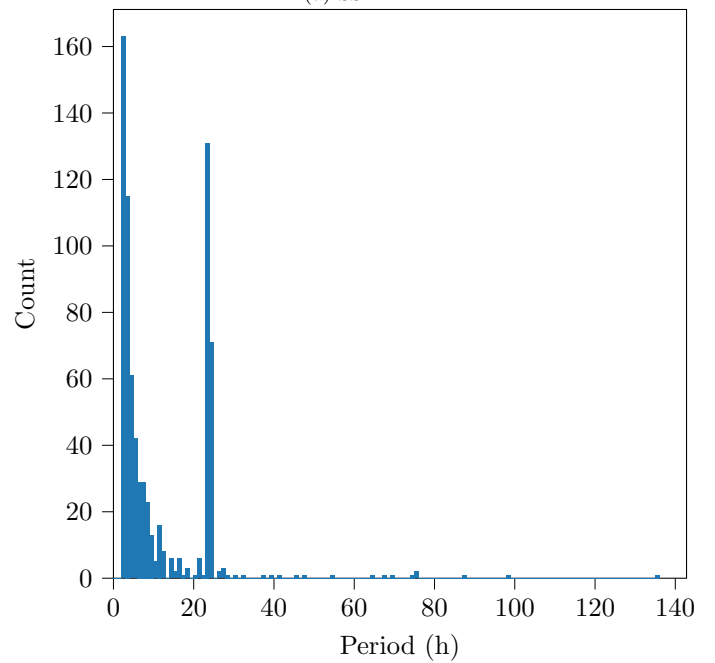}
    \caption{\ztf}
    \end{subfigure}
    \caption{The derived period distribution for the given survey using the Monte Carlo method.}
    \label{fig:mc_dist}
\end{figure}

\begin{figure}[H]
    \centering
    \begin{subfigure}[b]{.5\textwidth}

    \includegraphics[width=\textwidth]{figs/lsst/lsst_window.tikz}

    \caption{\lsst}
    \end{subfigure}
    \begin{subfigure}[b]{.5\textwidth}
  
    \includegraphics[width=\textwidth]{figs/ztf/ztf_window.tikz}

    \caption{\ztf}
    \end{subfigure}
    \caption{The derived period distribution for the given survey using the window method.}
    \label{fig:window_dist}
\end{figure}

\begin{figure}[H]
    \centering
    \begin{subfigure}[b]{.5\textwidth}
    \includegraphics[width=\textwidth]{figs/lsst/lsst_vdp.tikz}
    \caption{\lsst}
    \end{subfigure}
    \begin{subfigure}[b]{.5\textwidth}
   
    \includegraphics[width=\textwidth]{figs/ztf/ztf_vdp.tikz}

    \caption{\ztf}
    \end{subfigure}
    \caption{The derived period distribution for the given survey using the VanderPlas method.}
    \label{fig:vdp_dist}
\end{figure}

\end{appendices}

\end{document}